\pdfoutput=1
\documentclass{article}

     \usepackage[preprint]{neurips_2024}

\usepackage[utf8]{inputenc} % allow utf-8 input
\usepackage[T1]{fontenc}    % use 8-bit T1 fonts
\usepackage{hyperref}       % hyperlinks
\usepackage{url}            % simple URL typesetting
\usepackage{booktabs}       % professional-quality tables
\usepackage{amsfonts}       % blackboard math symbols
\usepackage{nicefrac}       % compact symbols for 1/2, etc.
\usepackage{microtype}      % microtypography
\usepackage{xcolor}         % colors

\usepackage{graphicx}
\usepackage{amsmath}
\usepackage{amssymb}
\usepackage{subfig}

\usepackage{threeparttable}
\usepackage{multirow}
\usepackage{amsmath}
\usepackage{caption}

\usepackage{algorithm}
\usepackage{algorithmic}

\usepackage{color}
\usepackage{bbding}
\usepackage{threeparttable}
\usepackage{dsfont}

\usepackage{wrapfig}
\usepackage{enumitem}

\title{MuseBarControl: Enhancing Fine-Grained Control in Symbolic Music Generation through Pre-Training and Counterfactual Loss}

\author{%
  Yangyang Shu\thanks{Project page: \url{https://ganperf.github.io/musebarcontrol.github.io/musebarcontrol/}}, Haiming Xu, Ziqin Zhou, Anton van den Hengel, Lingqiao Liu\thanks{Lingqiao Liu is the corresponding author.} \\
  School of Computer Science, The University of Adelaide\\
}

\begin{document}

\maketitle

\begin{abstract}
Automatically generating symbolic music—music scores tailored to specific human needs—can be highly beneficial for musicians and enthusiasts. Recent studies have shown promising results using extensive datasets and advanced transformer architectures. However, these state-of-the-art models generally offer only basic control over aspects like tempo and style for the entire composition, lacking the ability to manage finer details, such as control at the level of individual bars. While fine-tuning a pre-trained symbolic music generation model might seem like a straightforward method for achieving this finer control, our research indicates challenges in this approach. The model often fails to respond adequately to new, fine-grained bar-level control signals. To address this, we propose two innovative solutions. First, we introduce a pre-training task designed to link control signals directly with corresponding musical tokens, which helps in achieving a more effective initialization for subsequent fine-tuning. Second, we implement a novel counterfactual loss that promotes better alignment between the generated music and the control prompts. Together, these techniques significantly enhance our ability to control music generation at the bar level, showing a 13.06\% improvement over conventional methods. Our subjective evaluations also confirm that this enhanced control does not compromise the musical quality of the original pre-trained generative model.
\end{abstract}

\section{Introduction}

Symbolic music generation, which focuses on automatically creating music scores, has garnered increasing interest in recent years due to its intuitive readability and excellent editability~\cite{zhang2020butter,wu2022exploring}. Noteworthy advancements, such as Music Transformer \cite{anna2018music}, Museformer~\cite{yu2022museformer} and MuseCoco~\cite{lu2023musecoco}, have captivated both researchers and enthusiasts. Utilizing extensive datasets and sophisticated transformer architectures, these developments not only generate highly valuable music scores but also facilitate easy modification and editing, thanks to the accessibility provided by the score format.

Significant progress has been made in symbolic music generation, but there are still important limitations to acknowledge. One major challenge is the granular control over the music produced. Previous studies ~\cite{wu2022exploring,openai2023gpt,lu2023musecoco} have primarily focused on generating music using broad, overarching descriptions, allowing limited manipulation of elements like tempo and style across entire compositions. This lack of fine-grained control at the level of individual bars restricts the detailed alteration of musical elements. The ability to manage music at the bar level would be advantageous, offering users greater creative freedom and also enhancing applications in automatic music composition. For instance, attributes from a specific bar could be extracted and applied to generate another piece, facilitating style imitation. Bar-level control could improve the alignment of lyrics and melody, ensuring that the music accurately reflects the emotional cues of the lyrics. Furthermore, attributes from favoured music pieces can be identified and utilized to create new pieces using bar-level control, allowing for customized musical creation.

An effective method for implementing bar-level control is to fine-tune a foundational model with newly introduced control signals. This approach is particularly valuable due to the wide range of necessary controls that are difficult to fully anticipate during the initial training of the foundational model. The capability of adapt a trained model to new control is crucial as it enables the integration of diverse and unexpected controls, enhancing the model's utility without the need for complete retraining. Specifically, we can utilize bar-level music attributes extracted from the training set's music scores as prefix control prompts to train an autoregressive model, with the objective of optimizing the likelihood of the training samples. However, we've found that models trained in this manner often fail to adhere to the guidance of the bar-level attributes. Our analysis suggests that the model struggles with interpreting the meanings of these new control prompts, leading to music that does not accurately align with these prompts. Furthermore, when the training data is limited, the model is prone to overfitting, focusing more on minimizing loss rather than effectively using the control prompt to steer music generation.

%However, since the foundation model employs a transformer-based architecture and generates music sequences in an autoregressive manner, it may tend to rely on previously generated music tokens rather than the prefix prompts. This reliance significantly limits the model's controllability. Our experiments, illustrated in Figure~\ref{exp:tryk}, support this hypothesis. 

%A straightforward solution to achieve bar-level control is to fine-tune the existing foundation model to generate music based on the bar-level attribute available for each bar. Specifically, we could use bar-level music attributes as prefix prompts to guide music generation. However, since the foundation model employs a transformer-based architecture and generates music sequences in an autoregressive manner, it may tend to rely on previously generated music tokens rather than the prefix prompts. This reliance significantly limits the model's controllability. Our experiments, illustrated in Figure~\ref{exp:tryk}, support this hypothesis. 

To address this limitation, we propose two strategies to improve bar-level controllability. The first strategy involves pretraining the control prompt and fine-tuning the model on an auxiliary task designed to promote accurate alignment between the control prompts and the generated music tokens. The second strategy introduces a counterfactual loss that penalizes the model for neglecting the bar-level guidance. By implementing these two techniques together, we significantly enhance the accuracy of bar-level control without compromising the quality of the music produced.

%we introduce MuseBarControl, a groundbreaking foundation model that is specifically engineered to generate music based on bar-level instructions related to musical attributes. 
%\textcolor{red}{We can add more details about the motivation of our method in the above paragraph.} MuseBarControl addresses the limitation using two steps: pre-training task and counterfactual loss. The pre-training task is designed to directly associate control signals with corresponding musical tokens, facilitating a more effective initialization for subsequent fine-tuning phases. Counterfactual loss is introduced to enhance the alignment between the generated music and the control prompts. MuseBarControl constitutes a significant breakthrough in the field, granting composers, musicians, and producers extraordinary control over the music creation process. By enabling the specification of attributes such as chords on a bar-by-bar basis, MuseBarControl unlocks new avenues for producing music that precisely matches the creator’s artistic vision.

In sum, the key contributions of this work are outlined as follows:
\begin{itemize}[leftmargin=*]
    \item We conducted the first study in achieving fine-grained control of symbolic music generation based on the existing foundation model.
    %\item We  MuseBarControl, the first system designed to control symbolic music generation at the level of individual bars.
    \item We propose two innovative strategies, auxiliary task pre-training and counterfactual loss, to improve bar-level control in the foundation model. 
    %\item Musicality has been maintained (43.75\% vs 40.63\%), and the controllability of music generation at the bar level has been significantly enhanced by 13.06\%. This enhancement enables MuseBarControl to achieve highly efficient generation in practical applications.
    %\item MuseBarControl creates new opportunities for crafting music that closely matches the creator's vision, such as assisting composers, musicians, and producers in selecting the appropriate chords for individual bars during the music creation process.
\end{itemize}

\section{Related Work}
\label{related_work}
\subsection{Text-Driven Music Generation}
Text-driven music generation, aimed at creating high-quality music from textual descriptions, has attracted considerable attention from researchers due to its user-friendly editing capabilities. However, the scarcity of paired text-music data presents a major challenge. To address this, some studies have employed diffusion-based methods~\cite{zhu2023ernie, schneider2023mo} with self-collected text-audio datasets to facilitate text-audio music generation. MuLan~\cite{huang2022mulan} tackles the issue of data scarcity. They use techniques similar to those in CLIP~\cite{radford2021learning} to contrastively embed two modalities: music pieces and their textual annotations. Building on this, MusicLM~\cite{agostinelli2023musiclm} generates audio from MuLan's embeddings~\cite{huang2022mulan}, enabling text-to-music conversion without the need for paired data. However, MusicLM's process of sampling to acquire fine-grained acoustic tokens is computationally intensive. Other efforts, like MeLoDy~\cite{lam2024efficient}, seek to simplify music generation by efficiently translating conditioning tokens into sound waves. Furthermore, MusicGen~\cite{copet2024simple} introduces a single-stage transformer LM framework that models multiple streams of acoustic tokens in parallel. Despite significant advancements in text-driven music generation, the methods are still relatively crude, limiting users' ability to edit musical elements within the generated audio. The controllability and editability of the outputs remain constrained.

\subsection{Symbolic Music Generation}
Compared to text-driven music generation, symbolic music provides easier editing capabilities, allowing users to manipulate specific musical elements more effectively. The development of solutions for this task has evolved significantly, from grammar rules-based methods \cite{collins2009musical, garcia2011automatic, fernandez2013ai} to probabilistic models and evolutionary computation \cite{yanchenko2017classical, liu2016computational}, and more recently to neural networks and deep learning \cite{ycart2017study, briot2017deep, dong2018musegan}. The advent of transformer-based models, known for their successes in text generation \cite{li2022pretrained, ren2023you}, has also influenced music generation. The Music Transformer \cite{anna2018music} utilizes transformers with relative attention, proving highly effective for generating symbolic music. Museformer \cite{yu2022museformer} addresses challenges in long-sequence and music structure modelling by capturing music structure-related correlations, thus enhancing music generation efficiency. The recently introduced MuseCoco \cite{lu2023musecoco} offers precise control over symbolic music generation through specific attributes, using these attributes as a bridge to transition from text-to-music to attribute-to-music generation. MuseCoco enables the adjustment of various musical attributes, offering a level of control akin to the compositional process. However, this control is limited to entire compositions, diverging from a composer's typical approach, which often involves more granular control, such as bar-level manipulation. 
%Therefore, the development of a Granular-Controlled Music Score Generation system, capable of bar-level control and generation, is necessary to better align with traditional composition practices.

\section{Preliminaries}
\begin{figure}[htbp]
	\centering
	\includegraphics[width=4.5in]
 {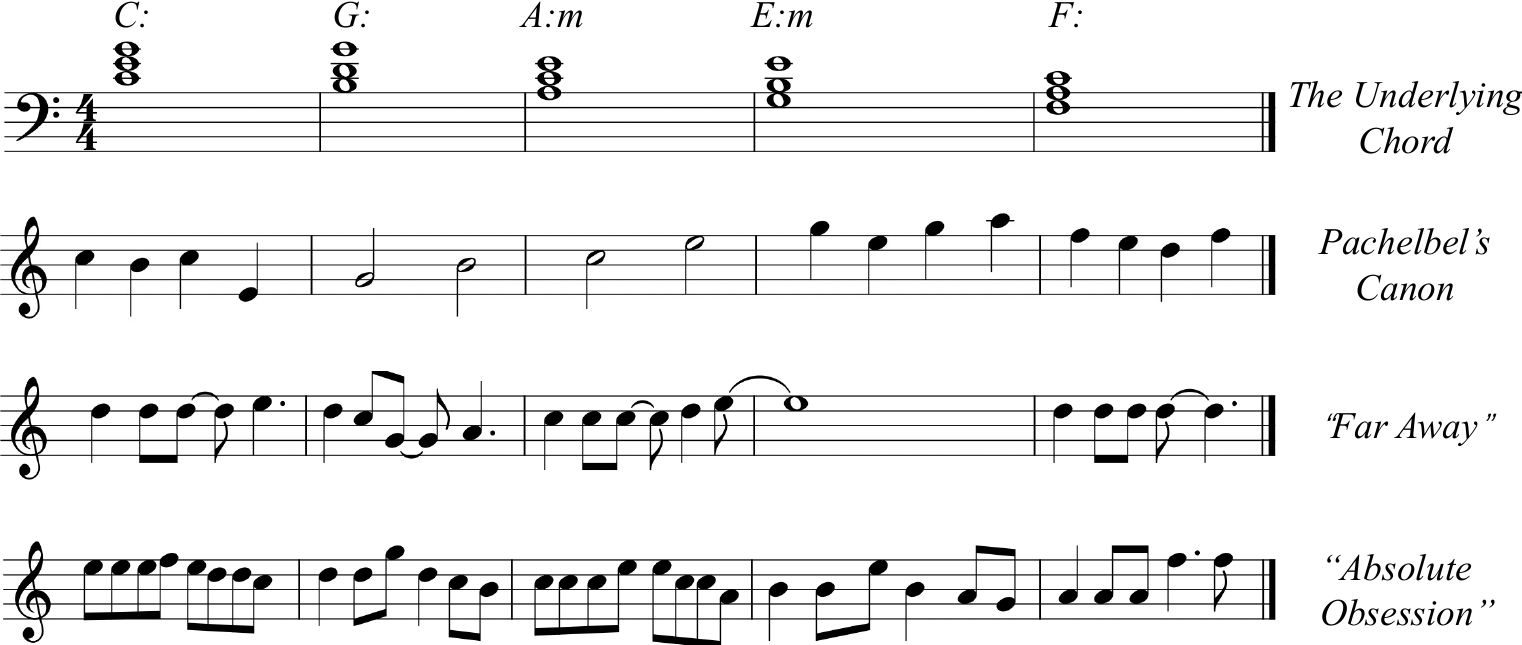}
	\caption{An example of three pop songs sharing the same chord progression. The top row displays five-column chords, while the bottom three rows represent the pop songs "Canon" by Pachelbel, "Far Away" by Jay Chou, and "Absolute Obsession" by Sam Lee.}
	\label{method:example}
\end{figure}
We begin by examining the MuseCoco model~\cite{lu2023musecoco}, which serves as the foundation for our method. MuseCoco is a text-to-music generative model that initially converts text instructions into a set of music attributes and then generates music tokens based on these attributes. Our approach builds upon this attribute-to-music generation model.
%structured in two distinct phases: 1) text-to-attribute understanding and 2) attribute-to-music generation. Our focus is primarily on elaborating and summarizing the methodologies employed in the latter phase.

In this model, a series of prefix tokens $\mathbf{x}=[x_1, x_2, ..., x_m]$ encodes the music attributes. Subsequently, the model generates a sequence of music tokens $\mathbf{y}=[y_1, y_2, ..., y_n]$. For additional details on the tokenization design, we refer to~\cite{lu2023musecoco}. The model is trained to maximize the log-likelihood of the ground-truth music sequences, as in the standard autoregressive model, namely:
\begin{align}\label{eq: musecoco}
    \mathcal{L} = -\sum_{i=1}^n  \log p(y_i|y_{<i}, x_{1:m} ), 
\end{align} where $y_{<i}$ indicates historic tokens before $i$.

One limitation of the MuseCoco model \cite{lu2023musecoco} is its focus solely on global music attributes, which describe the overall composition without supporting finer control, such as at the bar level. Fine-grained control is particularly valuable to both musicians and amateurs as it enables users to define specific properties for smaller segments of music. For instance, users can specify the chords\footnote{Chords are groups of notes played together that act as the foundation of music. They generate harmony and shape the emotional atmosphere of a piece. A sequence of changing chords, known as a chord progression, provides music with its rhythm and supports the melody. Often in music, especially in piano compositions, chords manifest not through simultaneous notes but through sequentially played notes, typically with the left hand, known as a broken chord. The specific chords used in a piece can be identified by analyzing the notes that appear throughout the composition.} in each bar and provide a chord progression, and then explore various musical outputs based on that progression. Figure~\ref{method:example} illustrates how different music pieces can share the same chord progression. Beyond its utility for human users, fine-grained bar-level control is also advantageous for automated composition. For example, bar-level music attributes from one piece could be used to conditionally regenerate another piece, facilitating style mimicry. Additionally, this level of control can help achieve a closer alignment between the emotional content of lyrics and the corresponding melody, such as matching an intensification of emotion in the lyrics with an ascending melody line.
%to prompt the generation of a music sequence $\mathbf{y}=[y_1, y_2, ..., y_n]$. Here, $m$ represents the number of musical attributes, and $n$ denotes the length of the resulting music sequence. These prefix tokens encapsulate the musical attributes, which directly influence the generation process. Consequently, the training loss for this phase can be expressed as follows:
%\begin{align}\label{eq: musecoco}
%    \mathcal{L} = -\sum_{i=1}^n \mathds{1}(q_i=y_i) \log \delta(p(q_i|y<i, x_1, x_2, ..., x_m )), 
%\end{align} where $\delta$ is the soft-max operation and $\mathds{1}(q_i=y_i)=1$ if $q_i=y_i$.

%\textcolor{red}{Add one paragraph to explain why we want to achieve bar-level control. Give example if possible}
%However, the music generated by MuseCoco does not allow for bar-level editing. As illustrated in Figure~\ref{method:example}, in the music score shown in the top part of the Figure, if we wish to change the chord in the first bar from G to C, or if we are dissatisfied with the A:m chord in the 6-th bar and want to replace it with a different A:m chord (from the top music score to bottom music score), such modifications are not feasible with MuseCoco. To address this, we aim to develop a bar-editable symbolic music generation system that allows for such detailed modifications.
\section{Method}
To attain precise control, we present MuseBarControl. This model overcomes the constraints of existing music score generation models that largely produce music based on broad and vague descriptions. Our method consists of three components: (1) we refine the control prompt in MuseCoco~\cite{lu2023musecoco} to facilitate bar-level control instead of global control. (2) we incorporate an auxiliary task to pre-condition the model and the newly implemented control prompts. (3) we introduce a counterfactual loss to enhance the adherence of the generated music to the bar-level control prompts.

\subsection{Control Prompt Augmentation}
%To enhance musical fluidity and achieve granular control, we introduce MuseBarControl. This model addresses the limitations of current music score generation models, which predominantly generate music based on broad and imprecise descriptions. MuseBarControl extends the concept utilized in Museoco~\cite{lu2023musecoco}, incorporating not only global musical attributes that define the overall character of the music but also specific attributes for each bar, such as chords, to facilitate a more detailed level of control. Essentially, MuseBarControl adopts the original Linear Transformer architecture from~\cite{katharopoulos2020transformers}, merging global and bar-level attributes to effectively address the challenges in Granular-Controlled Music Score Generation.
MuseCoco~\cite{lu2023musecoco} only incorporates global musical attributes that define the overall character of the music, which cannot achieve fine-grained bar-level control of music generation. To facilitate the latter, we first introduce a scheme to specify the bar-level music attribute. Specifically, MuseBarControl processes a sequence of music attribute tokens $X = X_g, X_1, X_2,...X_b$ as input, where $X_g=x_{g,1}, x_{g,2},...x_{g,|X_g|}$ comprises $|X_g|$ tokens representing global attributes, and $b$ denotes the number of music bars. Each $i$th bar $X_i=x_{i,1},x_{i,2},...,x_{i,|X_i|} $ contains $|X_i|$ tokens representing the attributes at the bar level, e.g., token $24$ could indicates the chord of the current bar is ``E:b'' (see Table~\ref{appendix:att_info} in Appendix). Those attributes can be extracted from the training music scores, i.e., from $y_1, y_2,..., y_n$. For example, there are existing algorithms~\footnote{We use the algorithm provided by \url{https://github.com/Rainbow-Dreamer/musicpy}.} to extract the chord implies in a given bar based on the distribution of the note pitches in the bar.
Position embeddings are added to the bar-level tokens within this sequence $X_1, X_2,...X_b$, distinguishing each bar's tokens shown in Figure~\ref{method:step2}. Tokens corresponding to the same bar are assigned identical position embeddings. Conversely, position embeddings are not used for the global tokens $X_g$. As a result, the input sequence in our method is encoded as follows:
\begin{align}\label{eq: input_sequence}
    X_g, X_1, X_2, ..., X_b, [SEP], y_1, y_2,..., y_n. 
\end{align}

A straightforward approach involves fine-tuning the MuseCoco model by optimizing the likelihood of the ground-truth music sequence given the bar-level attributes derived from that same sequence, say: 
%Specifically, we input a set of prefix tokens $X = X_g, X_1, X_2,...X_b$ which integrate global attributes and bar-level attributes. For each resulting music sequence $\mathbf{y}=[y_1, y_2, ..., y_n]$, we then optimize the following objective:
\begin{align}\label{eq: bft}
    \mathcal{L_{BFT}} = -\sum_{i=1}^n  \log p(y_i|y_{<i}, X), 
\end{align}where $\mathcal{L_{BFT}}$ denotes the \textbf{B}ar-level \textbf{F}ine-\textbf{T}uning loss.
%\hm{$\delta$ may be omitted as in eq.(\ref{eq: musecoco}) since $p(q_i|...)$ denotes the probability. and add a $\frac{1}{n}$ for averaging}
In practice, this process can be easily implemented via efficient parameter fine-tuning, such as LoRA \cite{hu2021lora}. However, as detailed in the experimental section (see Table \ref{exp:comp_w_musecoco}), this method proves to be less effective.
%\hm{may ref which table here} 
We find that the model tends to overlook the newly introduced control prompts in its efforts to maximize the likelihood. Consequently, although the loss decreases, the controllability of the new prompts does not improve. We hypothesize that this issue arises because the new control prompts are randomly initialized and they have not been supported by the foundation.
%\hm{why randomly init?}, 
As a result, the MuseCoco model struggles to adequately translate these controls into the music generation process. Furthermore, the model can more readily maximize likelihood by overfitting the data, such as by memorizing the training pieces, thereby diminishing any incentive for the network to adhere to the control signals.

To address this issue, this paper proposes two strategies to improve the controllability of the network. 

%Our approach involves sequentially applying two steps during the training phase, as detailed in Sections~\ref{method:PA} and \ref{method:intervention}, with the inference process described in Section \ref{method:inference}.
\subsection{Control Prompts Pre-Adaptation via an Auxiliary Task (PA)}
\label{method:PA}
\begin{figure}[t] 
 % \begin{minipage}[b]{0.485\textwidth}
 \parbox{.48\linewidth}{
  % \end{minipage}% 
  % \begin{minipage}[b]{0.485\textwidth} 
    \centering 
    \includegraphics[width=0.48\textwidth]{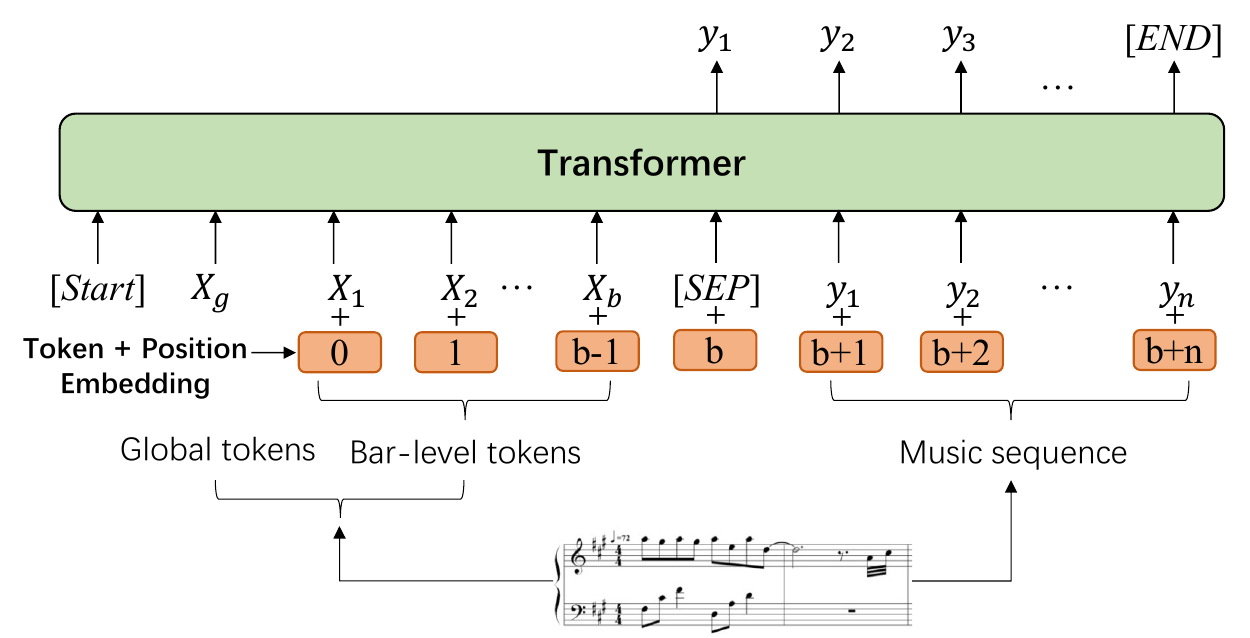} 
    %\vspace{0.9em}
    \caption{Control prompt augmentation.} 
    \label{method:step2} 
    }
    \hfill
    \parbox{.48\linewidth}{
    \centering 
    \includegraphics[width=0.48\textwidth]{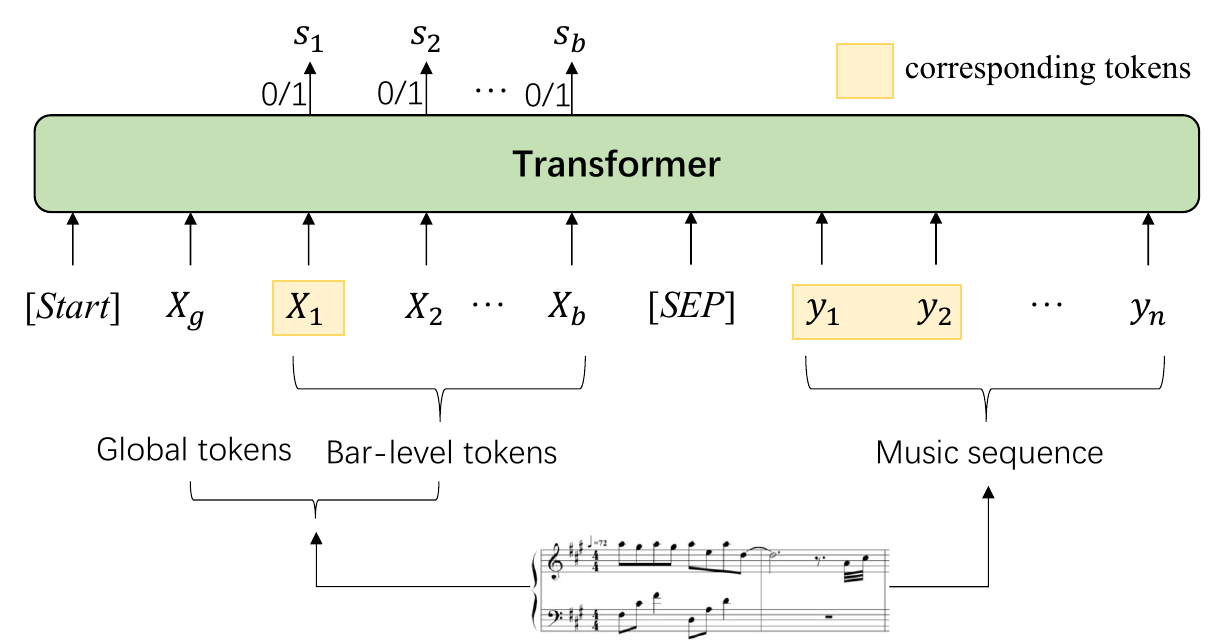} 
    % \ç{r1_r2.png} 
    
    \caption{Pre-adapt the bar-level control.} 
    \label{method:step1} 
    }
    
  % \end{minipage}% 
\end{figure}
The first strategy is to utilize an auxiliary task to pre-adapt the bar-level control mechanism of the MuseCoco model. In this context, "pre-adapt" refers to modifying the embeddings of the control prompts and the LoRA parameters. The auxiliary task is framed to meet two key criteria: firstly, it must be closely related to the task of bar-level controlled generation, ensuring that the parameters refined during the auxiliary task can be effectively transferred to the main generation task. Secondly, the model can only accomplish the auxiliary task by genuinely utilizing the interactions between the control prompts and the corresponding music tokens. Specifically, to achieve bar-level controlled generation, it is essential for the model to comprehend which music tokens and specific attributes of those tokens are governed by a bar-level prompt. Accordingly, we propose a recognition task: the model is presented with a sequence of bar-level control prompts alongside a sequence of music tokens, and it must determine whether the music sequence conforms to the guidelines set by the bar-level control prompts. For this purpose, we develop new linear projection heads that are trained on the output embeddings of the bar-level control prompts. These projections perform a binary classification task to assess whether the specified bar-level control, as indicated by the control prompt, has been adhered to in the music sequence. This setup is depicted in Figure~\ref{method:step1}. Successfully completing this task hinges on the accurate correlation between the control prompts and the music tokens, a skill that is directly transferable to controlled music generation.

%The fundamental concept of PA is to refine the training approach beyond the typical MuseCoco style by ensuring that the model recognizes the specific bar to which each music sequence corresponds. This step is designed to enable the model to discern the relationships between bar-level attributes and the corresponding music tokens within each bar. To accomplish this, we train new linear projections tasked with predicting whether the bar-level attributes are accurately aligned with their respective bars. Figure~\ref{method:step1} illustrates this process.

In our implementation, we randomly select the number of $t$ bars, denoted as $\mathcal{M}$ and $|\mathcal{M}| = t$, and modify the corresponding bar-level tokens in the input to reflect that the attributes of these bars do not match the expected values. Then we create a sequence of ground-truth prediction labels $\{s_i\}$, with $s_i = 1$ for unmodified bar and $s_i = 0$ for modified bars (``1'' indicates ``match'' and ``0'' indicates ``unmatch''). The training loss for this step is expressed as follows:
\begin{align}\label{eq: pa}
    \mathcal{L_{PA}} = -\sum_{i\in \mathcal{M} }\log (1-p(s_i|\mathbf{y}, X')) - \sum_{j \not \in \mathcal{M}} \log(p(s_j|\mathbf{y}, X')),
\end{align} 
%where $t$ represents the number of selected bars\hm{$t$ does not in eq(4)}, and 
where $X'$ represents the set of all bar-level attribute tokens, with modifications made to some bars,  such as changing the chord from C: to A:m.

\subsection{Improving Controllability via Counterfactual Loss (CF)}
\label{method:intervention}
%In MuseCoco~\cite{lu2023musecoco}, various musical attributes are combined and sequenced to facilitate comprehensive musical expressions and transformations. However, achieving granular control at the bar level remains challenging. To make the model easier to control and understand how the attributes are influencing bar-level music, we introduce a modification to the MuseCoco method, MuseCoco with Bar-level Training (BFT). This adaptation involves training with both global and bar-level attribute tokens to steer music generation. The process is depicted in Figure~\ref{method:step2}.

%Specifically, we input a set of prefix tokens $X = X_g, X_1, X_2,...X_b$ which integrate global attributes and bar-level attributes. For each resulting music sequence $\mathbf{y}=[y_1, y_2, ..., y_n]$, we then optimize the following objective:
%\begin{align}\label{eq: musecoco}
%    \mathcal{L_{BFT}} = -\sum_{i=1}^n \mathds{1}(q_i=y_i) \log \delta(p(q_i|y<i, X)), 
%\end{align} where $\delta$ is the soft-max operation and $\mathds{1}(q_i=y_i)=1$ if $q_i=y_i$.
The second strategy incorporates the use of a counterfactual loss to verify that the generated tokens are genuinely influenced by the bar-level control prompts. Our rationale is that if the music tokens of a particular bar are truly driven by its associated bar-level control prompt, then altering the control prompt should result in a substantial decrease in the likelihood of those specific music tokens. Specifically, we randomly replace the bar-level attributes $X_i\in \{X_1,X_2,...,X_b\}$ with a different value within the attribute, denoted as $\overline{X_i}$. We then measure the change of the negative log-likelihood of the $i$th bar's token, represented by the difference $\mathcal{J}_2-\mathcal{J}_1$, where $\mathcal{J}_1 = -\frac{1}{|bar_i|}\sum_{i\in bar_i} \log p(y_i|y_{<i}, X_g, X_1,..., X_i,..., X_b)$ is the negative log-likelihood before the change in this bar, and $\mathcal{J}_2 = -\frac{1}{|bar_i|}\sum_{i\in bar_i} \log p(y_i|y_{<i}, X_g, X_1,..., \overline{X_i},..., X_b)$ indicates the negative log-likelihood after the change in this bar. %To maintain the new prompt distribution $X_g, X_1,..., \overline{X_i},..., X_b$ as close as possible to the original $X_g, X_1,..., X_i,..., X_b$~\cite{koh2017understanding}, 
%$\overline{X_i}$ can be an arbitrarily sampled attribute token. In our implementation, we set it to the same value as $X_{i-1}$ when $X_i\neq X_{i-1}$ to further promote the model's awareness of the bar-level attribute's governance range.  
 $\overline{X_i}$ represents a randomly selected attribute token. In our implementation, to enhance the model's recognition of the governance range of bar-level attributes, we assign $\overline{X_i}$ the same value as $X_{i-1}$ whenever $X_i\neq X_{i-1}$. 
 
 This approach is designed to reinforce the model's awareness of the influence exerted by bar-level attributes. The counterfactual loss is thus defined as:
\begin{align}\label{eq: cf}
    \mathcal{L_{CF}} = max\{0, \eta - (\mathcal{J}_2 - \mathcal{J}_1)\}, 
\end{align} where the counterfactual loss $\mathcal{L_{CF}}$ is designed to promote a significant decrease in the log-likelihood when the alignment between the control prompt and the music tokens is disrupted after modifying $X_i$. $\eta$ is a margin parameter specifying the desired log-likelihood change. 
%\hm{$(\mathcal{J}_1 - \mathcal{J}_2)$ should be $|\mathcal{J}_1 - \mathcal{J}_2|$? otherwise the loss can be infinite large}\textcolor{blue}{Probably not. If $J_1$<$J_2$, there will be a large penalty!}
%By adjusting this margin, we can control the extent to which the model's predictions rely more on the prefix attributes than on previous tokens in an autoregressive manner.

Overall, our method first pre-adapts the model with the training task described in Section \ref{method:PA} and then apply the counterfactual loss together with the bar-level fine-tuning loss:
\begin{align}\label{eq: BFT_CF}
    \mathcal{L} =  \mathcal{L_{BFT}} + \lambda\mathcal{L_{CF}},
\end{align} where $\lambda$ is the trade-off hyperparameter.

\subsection{Inference}
\label{method:inference}
%During the inference stage, music sequences are generated on a bar-by-bar basis. Within each bar, we can attempt to sample up to $K$ times. If a sample correctly captures the intended bar-level attributes, we continue generating the sequence from the next bar onward. However, if the sample lacks the correct attributes, we proceed by selecting the token with the highest probability to continue the prediction. When $K=1$, the above process reduces to the standard auto-regressive model sampling procedure. Generally, larger $K$ leads to better controlability but lower sampling efficiency. Note that for many bars the model may generate valid music in the first trial. As a result, the actual sampling time may only go sublinearly with $K$

During the inference stage, music sequences are generated on a bar-by-bar basis. Within each bar, up to $K$ sampling attempts are allowed. If a sample accurately reflects the intended bar-level attributes, the generation continues from the subsequent bar. Conversely, if $K$ samples all fail to exhibit the correct attributes, the token with the highest probability is chosen to continue the sequence prediction. When $K=1$, this process simplifies to a typical auto-regressive model sampling procedure. Generally, a larger $K$ enhances controllability at the expense of reduced sampling efficiency. It is noteworthy that in many cases, the model may produce valid music on the first attempt, implying that the actual number of sampling operations may increase sublinearly with $K$.

%Formally, the inference stage can be described as follows:
%\begin{align}\label{eq: musecoco}
%    \{y_i\}_{i=1}^n = cat(try_K(y_i^j))_{j=1}^b,
%\end{align} where $y_i^j$ denotes the tokens $y_i$ generated in the $j$th bar, and $tryK$ represents the number of attempts to sample, up to $K$ times. Here, $tryK(y_i^j)$ is decoded into a musical expression that is expected to match the bar-level attributes specified by the prompt $X_j$.

\section{Experiments}
In this section, we assess our method through a case study focused on bar-level chord control. In other words, the bar-level attribute corresponds to the chord for each bar. This is useful as in music composition, particularly in pop music, it is customary to first establish a chord progression pattern before composing the music. It's important to note that the objective of our experiment is to evaluate the effectiveness of the proposed methods in enhancing control based on a pre-trained music model. We are not aiming to optimize performance specifically for chord-controlled music generation. 
\subsection{Experimental Setting}
\subsubsection{Datasets}
% \begin{wraptable}{r}{0.25\textwidth}
%     \caption{Statistics of Top 8 chord proportions (\%) in the POP909 dataset.}
% \label{exp:pop909_chord}
% \resizebox{\linewidth}{!}{
%     \begin{tabular}{cc}
% \toprule
%  Chord & Proportion \\ \midrule
%         A:m        & 18.31    \\
%         C:        & 17.11       \\
%         G:        & 14.13       \\
%         F:        & 9.94       \\
%         D:m        & 8.03       \\
%         E:m        & 6.97       \\
%         A:m7        & 3.91       \\
%         E:m7        & 3.24      \\\bottomrule
% \end{tabular} }
% \end{wraptable}

% \usepackage{multirow}

In our study, we use the POP909 dataset \cite{wang2020pop909}
%\footnote{There are two primary reasons for selecting the POP909 dataset as our training set: (1) Pop music typically features distinct chord progression patterns, making it ideal for this study. (2) Pop piano music is more accessible for general audiences to evaluate its quality. } 
to train and evaluate the proposed method. 
%(1) it consists of popular songs, making it easier for the audience to judge chord arrangements in each bar, and (2) pop songs have broad applications in music editing and composition. 
This dataset comprises multiple renditions of the piano arrangements for 909 popular songs, totalling approximately 60 hours of music. These arrangements are meticulously crafted by professional musicians and are provided in MIDI format. On average, each song consists of 270 bars. We chose this dataset for two reasons: (1) Pop music typically features distinct chord progression patterns, making it ideal for this study. (2) Pop piano music is more accessible for general audiences to evaluate its quality.  Following \cite{lu2023musecoco}, we randomly selected three 16-bar clips from each MIDI file. From these clips, global attributes are extracted from the entire clip, while bar-level attributes—chords, are extracted from each individual bar within the 16-bar length clip. The specific predefined musical attribute values, including global attributes sourced from the MuseCoco project \cite{lu2023musecoco} and chord attributes derived from MIDI files, are displayed in Appendix~\ref{Appendix_att}. Details regarding the distribution of chords within this dataset are provided in Appendix~\ref{Chord_Proportions}. To convert the MIDI files into token sequences, we employ a REMI-like representation method~\cite{huang2020pop}. For training, validation, and testing purposes, we partition the songs into three sets, with a split ratio of 8:1:1 for training, validation, and testing, respectively.

\subsubsection{Implementation details}
We employ the Linear Transformer architecture as our backbone model~\cite{katharopoulos2020transformers}, configured with a causal attention mechanism spanning 24 layers and utilizing 24 attention heads. The hidden size is set to 2048, while the feedforward network (FFN) hidden size is 8192. During the training phase, we began by initializing the MuseCoco-base weights~\cite{lu2023musecoco} with fp16 precision and subsequently applied a fine-tuning approach. Within the attention layers, LoRA Adapters~\cite{hu2021lora} was incorporated, with a rank size of $r=8$. The maximum sequence length was set to 5120. We use validation performance to set the margin $\eta$ to 0.05. To execute the fine-tuning process, we utilized 4 40GB-A100 GPUs, conducting 50 epochs for the first strategy (PA), and 40 epochs for the second strategy (BFT and CF). We utilize a batch size of 4 and employ the Adam optimizer~\cite{kingma2014adam} with a learning rate of $2 \times 10^{-4}$, incorporating a warm-up step of 16,000 and an invert-square-root decay schedule. For inference, we consider the top $15$ highest probabilities as potential prediction hypotheses and perform sampling $K = 15$ times.

\subsubsection{Compared models} 
In this investigation, we conduct a comparative analysis between our proposed method and \textbf{MuseCoco}~\cite{lu2023musecoco} for symbolic music generation. Since MuseCoco is only fed with global musical attributes, we examine MuseCoco with \textbf{B}ar-level \textbf{F}ine-\textbf{T}uning, represented as \textbf{BFT} which builds upon the MuseCoco model by incorporating bar-level training techniques. This approach involves aligning the music attributes with each bar chord in the input.%We specifically evaluate the generation capabilities of each method when provided with text descriptions that specify chord alignments for each bar.

%Specifically, we examine MuseCoco with Bar-level Training, which builds upon the MuseCoco model by incorporating bar-level training techniques. This approach involves aligning the music attributes with each bar chord in the input. Additionally, we evaluate GPT-4, focusing on its generation capabilities when provided with text descriptions where each bar chord is aligned.
%\begin{itemize}[leftmargin=*]
%    \item \textbf{GPT-4:}
%    \item \textbf{MuseCoco:} 
%\end{itemize}

\subsubsection{Objective evaluation}
\begin{itemize}[leftmargin=*]
    \item \textbf{Chord Accuracy:} We use chord accuracy to evaluate the alignment between the prompted chords and the chords generated during the symbolic music generation process. This measure provides insight into the bar-level controllability afforded by our proposed method.
    \item \textbf{Global Attribute Accuracy:} We use global attribute accuracy to evaluate the alignment between prompted global musical attributes and those generated during the inference process. This metric offers insight into the sample-level controllability of our method.
\end{itemize}

\subsubsection{Subjective evaluation}
\begin{itemize}[leftmargin=*]
    \item \textbf{Musicality:} measures the extent to which generated music resembles music created by humans.
\end{itemize}
We randomly selected 20 pieces of music generated by MuseCoco performed with piano and our model and created a survey. We then enlisted 16 piano teachers to evaluate which pieces resembled human-created music more closely. The survey options were `Music 1', `Music 2', or `Similar', where Music 1 and Music 2 were randomly assigned to either the MuseCoco generation or ours~\footnote{We have included the 20 pieces in the Supplementary Material and converted the MIDI files to MP3 format.}. The degree of musicality was gauged based on the percentage of votes received for each option. 
\subsection{Compared with MuseCoco}
\begin{table}
    \caption{Comparative analysis of objective and subjective evaluations among MuseCoco, BFT, and MuseBarControl.}
    \centering
\label{exp:comp_w_musecoco}
%\resizebox{\linewidth}{!}{
    \begin{tabular}{cccc}
\toprule
 Method & Musicality (\%) & Chord Acc. (\%) & Average Global Attribute Acc. (\%) \\ \midrule
        MuseCoco        & 40.63    & -  &78.14 \\
        BFT        & -   & 65.27  & 81.54   \\
        MuseBarControl        &  \textbf{43.75}    & \textbf{78.33 }  & \textbf{82.55}\\\bottomrule
\end{tabular} %}
\end{table}
Table~\ref{exp:comp_w_musecoco} presents the results of both objective and subjective evaluations. The findings are as follows: 1) In terms of musicality, MuseBarControl achieves performance comparable to MuseCoco, with scores of 43.75\% versus 40.63\% respectively (the full survey statistic is shown in Figure~\ref{exp:music_survey}.). This equivalence in scores demonstrates that MuseBarControl preserves the musicality inherent in MuseCoco. 2) Regarding chord accuracy, MuseCoco lacks the capability to generate specific chords aligned with each bar. Compared to BFT, MuseBarControl exhibits a significant improvement in chord accuracy by 13.06\%, underscoring its enhanced controllability at the bar level. 3) Concerning average global attribute accuracy, MuseBarControl outperforms BFT by a slight margin of 1.01\%, and both show a substantial improvement over MuseCoco. The data indicates that aligning music attributes with each bar chord in the input significantly enhances global attribute accuracy. This improvement suggests that the effective generation of bar-level attributes can positively influence global attribute generation.

\subsection{Ablation Study}
\subsubsection{Component-wise analysis}
\begin{table}
    \caption{Analyzing the impact of proposed components on chord accuracy. Results Based on $K=15$ sampling times.}
    \centering
\label{exp:component-wise}
%\resizebox{\linewidth}{!}{
    \begin{tabular}{cccc}
\toprule
 BFT &PA  & CF & Chord Accuracy (\%) \\ \midrule
    $\checkmark$      &  $\times$    & $\times$   & 65.27   \\
            $\checkmark$      &  $\checkmark$     & $\times$    & 72.22\\
            $\checkmark$         & $\times$     & $\checkmark$   &  71.15  \\
       $\checkmark$         &  $\checkmark$    & $\checkmark$   & \textbf{78.33}\\\bottomrule
\end{tabular} %}
\end{table}
In this section, we conduct ablation studies to evaluate the impact of each component within our method. The findings are detailed in Table~\ref{exp:component-wise}. From the table, it can be seen that employing BFT alone only yields a chord accuracy of 65.27\% during the inference stage. Notably, the inclusion of PA or CF significantly enhances performance, with improvements of 6.95\% and 5.88\%, respectively. This highlights the critical roles that PA and CF play in improving chord recognition and enhancing bar-level control. Furthermore, the simultaneous use of PA and CF leads to the best performance, achieving a substantial increase in chord accuracy from 72.22\% to 78.33\% with PA and from 71.15\% to 78.33\% with CF.
These components enhance chord accuracy from distinct perspectives: PA aligns bar-level attributes and musical sequences as input, helping the model to effectively capture the relationships between bar-level attributes and corresponding music sequences; 
%BFT uses global attributes and bar-level attributes as prefix prompts, assisting the model in establishing mappings from attribute tokens to music generation; 
CF is designed to ensure the model correctly responds to the prefix bar-level prompts and avoids the trivial solution of generating the next tokens solely based on previous tokens.
\subsubsection{Impact of \texorpdfstring{$K$ \textemdash}{Lg} the number of sampling in inference}

\begin{figure}[t] 
 % \begin{minipage}[b]{0.485\textwidth}
 \parbox{.44\linewidth}{
    \centering 
    \includegraphics[width=0.44\textwidth]{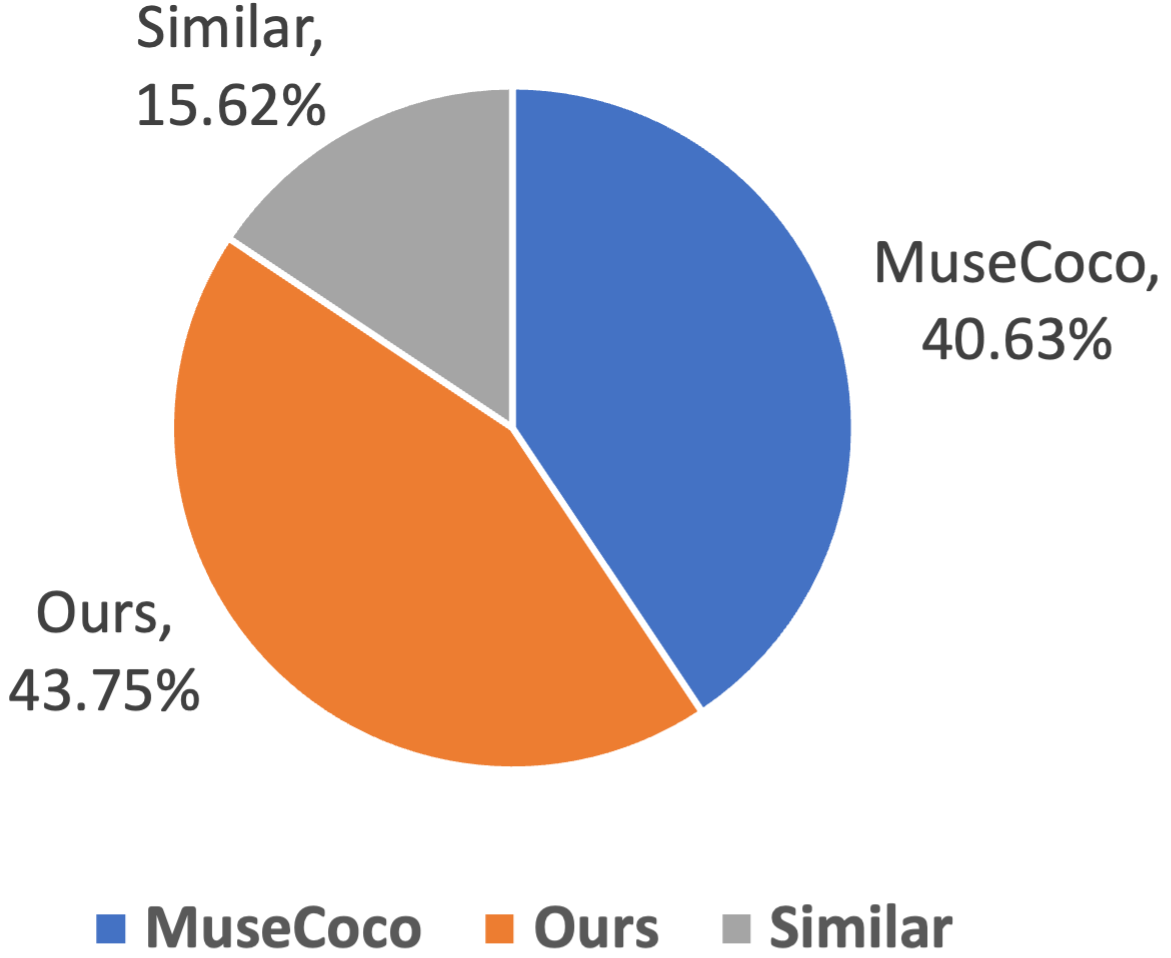} 

    \caption{The vote percentages of music generated by MuseCoco and our method, as judged by 16 piano teachers for similarity to human creation.} 
    \label{exp:music_survey} 
    }
    \hfill
    \parbox{.44\linewidth}{
  % \end{minipage}% 
  % \begin{minipage}[b]{0.485\textwidth} 
    \centering 
    \includegraphics[width=0.44\textwidth]{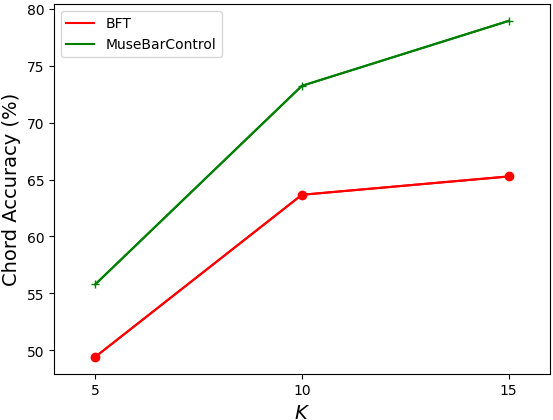} 
    %\vspace{1.6em}
    
    \caption{Chord accuracy of BFT and MuseBarControl w.r.t. the number of different $K$.} 
    \label{exp:tryk} 
    }
  % \end{minipage}% 
\end{figure}

% \begin{figure}[tbp]
% 	\centering
% 	\includegraphics[width=2.5in]{music_survey1.png}
% 	\caption{.}
% 	\label{exp:music_survey}
%  \end{figure}
 
% \begin{figure}[tbp]
% 	\centering
% 	\includegraphics[width=2.5in]{tryK.png}
% 	\caption{Chord accuracy of BFT and MuseBarControl w.r.t. the number of different $K$.}
% 	\label{exp:tryk}
%  \end{figure}

In our method, during the inference phase, we select music sequences bar-by-bar from the $K$ sampling attempts. This section explores the impact of the value of $K$. Figure~\ref{exp:tryk} displays the accuracy achieved with different $K$ values. The results indicate that as $K$, the number of sampling attempts, increases, chord accuracy rises significantly. Additionally, compared to BFT, MuseBarControl consistently delivers superior performance across various $K$ values, with the performance gap between BFT and MuseBarControl widening as $K$ increases.

\subsubsection{Impact of the parameter \texorpdfstring{$\lambda$}{Lg}}
\begin{table*}[t]\tiny
  % \begin{minipage}[b]{0.485\textwidth} 
  \parbox{.46\linewidth}{
    \centering
    \captionof{table}{Chord accuracy (\%) of the proposed method with different $\lambda$.}
    \resizebox{0.75\linewidth}{!}{
    \begin{tabular}{cc}
\toprule
 $\lambda$ & Chord Accuracy \\ \midrule
        0        & 72.22        \\
        1e1      & 73.79        \\
        1e2       & 76.42        \\
        1e3        & \textbf{78.33}    \\
        1e4       &76.64\\\bottomrule
\end{tabular} 
\label{exp:lamda} 
    }
    } % parbox
    \hfill
  % \end{minipage} ~~~
  % \begin{minipage}[b]{0.4\textwidth} 
  \parbox{.46\linewidth}{
    \centering
    \captionof{table}{The average inference time on MuseCoco and our method per sample. $K=5$ and $K=15$ denote the number of samples in inference. The inference times were measured on an NVIDIA RTX 4090 GPU.}
    \resizebox{0.8\linewidth}{!}{
     \begin{tabular}{cc}
\toprule
 Method & Runtime \\ \midrule
        MuseCoco        & 3 mins        \\
        Ours ($K=5$)       & 4 mins         \\
        Ours ($K=15$)        & 6 mins     \\\bottomrule
\end{tabular}
\label{exp:complexity} 
    }
    } % parbox
  % \end{minipage} ~~~
\end{table*}

% \begin{wraptable}{r}{0.25\textwidth}
%     \caption{Chord accuracy (\%) of the proposed method with different $\lambda$.}
% \label{exp:lamda}
% \resizebox{\linewidth}{!}{
%     \begin{tabular}{cc}
% \toprule
%  $\lambda$ & Chord Accuracy \\ \midrule
%         0        & 72.22        \\
%         1e1      & 73.79        \\
%         1e2       & 76.42        \\
%         1e3        & 78.33       \\
%         1e4       &76.64\\\bottomrule
% \end{tabular} }
% \end{wraptable}
To assess the impact of the parameter $\lambda$, we conducted experiments using varying values of $\lambda$. Table~\ref{exp:lamda} presents the chord accuracy corresponding to five different settings of $\lambda$. As observed, when $\lambda$ is set to 0, the approach defaults to PA+BFT. As $\lambda$ increases from 0 to 1e3, there is a gradual improvement in chord accuracy, reaching a peak of 78.33\% when $\lambda$ is approximately 1e3. Beyond this point, the performance begins to decline slightly.
\subsection{Complexity Analysis}
% \begin{wraptable}{r}{0.25\textwidth}
%     \caption{The average inference time per sample.}
% \label{exp:complexity}
% \resizebox{\linewidth}{!}{
%     \begin{tabular}{cc}
% \toprule
%  Method & time \\ \midrule
%         MuseCoco        & 3 mins        \\
%         Ours ($K=5$)       & 4 mins        \\
%         Ours ($K=15$)        & 6 mins      \\\bottomrule
% \end{tabular} }
% \end{wraptable}
Assuming the length of the music sequence is $n$ and the number of sampling attempts per bar is $K$, the time complexity of the inference step in MuseCoco is $O(n)$, whereas in our method it is $O(Kn)$. Although the time complexity scales with the number of sampling attempts $K$, the actual sampling often terminates earlier if the model successfully matches the chord early on. Consequently, the average number of samples taken at each bar is typically less than $K$. The average inference time per sample is presented in Table~\ref{exp:complexity}.
% \subsection{Case Analysis}
% \subsubsection{General case generation}
% \begin{figure}[htbp]
% 	\centering
% 	\includegraphics[width=4.5in]
%  {case_study1.pdf}
% 	\caption{ A segment of generated music. The model was provided with bar-level prompts from bar 1 to bar 16 as follows: D:m, D:m, D:m, E:m7, E:m7, E:m7, E:m7, E:m7, G:, G:, G:, A:m, A:m, A:m, A:m, A:m. As observed, the chords in the generated music perfectly align with the prompts.}
% 	\label{case_study}
% \end{figure}
% Figure~\ref{case_study} displays a segment of music generated by our method. As illustrated, bars 1 through 12 feature broken chords, starting with three D:m chords, transitioning to five E:m7 chords, and followed by three G: chords. From bars 13 to 16, the music shifts to four A:m column chords, enhancing the emotional depth of the piece. All chords in the generated music precisely match the prompts designated for each bar. For additional examples, including both successful and failed cases, please refer to Appendix~\ref{Appendix_pinao_rolls}.

\subsection{Comments from Musician and Composer}
We invited musicians to provide feedback on the music created by our system and composers to experience how our system aids the creative process. Their guidance and comments are as follows:

From Musician: ``\textit{I was truly impressed by the music produced by this system; its performance is remarkable. The quality of the music is very similar to that of human compositions, and some of the bar chord arrangements are astonishing}''.

From Composer: ``\textit{This system for chord arrangement and music creation significantly reduces my composition time. I found it to be a great source of inspiration. For instance, when I wanted to arrange the next chord as an A:m chord, the system provided many options to choose from. However, it uses a lot of broken chords, which composers typically don't use as frequently in their compositions. It would be beneficial if this aspect could be improved in the future}''.

\section{Limitation and Future Work} 
This work focuses on attribute-to-music generation, directly specifying attribute values to control the bar-level music generation process. However, this approach may not be user-friendly for those who prefer to use text descriptions for control. Therefore, in the future, we aim to develop a more user-friendly interface that allows bar-level music generation from text descriptions, enabling users to create and edit bars using natural language. Also, we plan to build a system incorporating more diverse control signals.

\section{Conclusion}
In this paper, we introduce MuseBarControl, a method that allows for finer detail control at the level of individual bars, significantly advancing the field of automated music score composition and alignment. This innovative approach offers substantial value and potential for both musicians and amateurs, enhancing creative efficiency and providing greater control over the composition process. Our solution contains two innovative strategies that enhance bar-level control without compromising the quality of the music produced. These designs enable the model to accurately adapt to the adjusted bar-level attributes as new control prompts, thereby achieving impressive bar-level controllability. 

Our research demonstrates the feasibility of bar-level editing in AI technologies. Successful results in chord control and generation, as shown in Appendix~\ref{Appendix_pinao_rolls}, suggest that more bar-level attributes, such as melody trends, can be explored in the future. We hope that further inspiration for bar-level attributes will enhance the ability to edit and control bar generation, thereby improving music creation. It is possible to adopt MuseBarControl for more bar-level attribute edits like melody trends. We hope MuseBarControl will be even more useful with additional bar-level attributes, utilizing the proposed two training strategies.

{\small
\bibliographystyle{abbrv}
\bibliography{main}
}

\appendix

\section{Appendix}
\subsection{Pre-defined Musical Attributes}
\label{Appendix_att} 
Table~\ref{appendix:att_info} displays the global attributes and their values, as well as the bar-level chord attributes and their values.
\begin{table}[htbp]
\caption{Global and bar-level attribute value.``NA" indicates that the attribute is not mentioned in the text, while ``N.C." denotes the absence of any chords.}
\label{appendix:att_info}
\begin{tabular}{lll}
\hline
Type                    & Attribute        & Values                                                                                                                                                                                                                                                                                                                                                                                                                                                                                                                                                                                                                                                                                                                                                                                                                                                                                                                                                                                                                                                                                                                                                  \\ \hline
\multirow{8}{*}{Global} & Instrument       & piano. 0: played, 1: not played, 2:NA                                                                                                                                                                                                                                                                                                                                                                                                                                                                                                                                                                                                                                                                                                                                                                                                                                                                                                                                                                                                                                                                                                                   \\
                        & Pitch Range      & 0-11: octaves, 12:NA                                                                                                                                                                                                                                                                                                                                                                                                                                                                                                                                                                                                                                                                                                                                                                                                                                                                                                                                                                                                                                                                                                                                    \\
                        & Rhythm Intensity & 0: serene, 1: moderate, 2: intense, 3: NA                                                                                                                                                                                                                                                                                                                                                                                                                                                                                                                                                                                                                                                                                                                                                                                                                                                                                                                                                                                                                                                                                                               \\
                        & Bar              & 0: 1-4 bars, 1: 5-8 bars, 2: 9-12 bars, 3: 13-16 bars, 4: NA                                                                                                                                                                                                                                                                                                                                                                                                                                                                                                                                                                                                                                                                                                                                                                                                                                                                                                                                                                                                                                                                                            \\
                        & Time Signature   & 0: 4/4, 1: 2/4, 2: 3/4, 3: 1/4, 4: 6/8, 5: 3/8, 6: others, 7: NA                                                                                                                                                                                                                                                                                                                                                                                                                                                                                                                                                                                                                                                                                                                                                                                                                                                                                                                                                                                                                                                                                  \\
                        & Key              & 0: major, 1: minor, 2: NA                                                                                                                                                                                                                                                                                                                                                                                                                                                                                                                                                                                                                                                                                                                                                                                                                                                                                                                                                                                                                                                                                                                               \\
                        & Tempo            & \begin{tabular}[c]{@{}l@{}}0: slow (\textless{}=76 BPM), 1: moderato (76-120 BPM),\\  2: fast (\textgreater{}=120 BPM), 3: NA\end{tabular}                                                                                                                                                                                                                                                                                                                                                                                                                                                                                                                                                                                                                                                                                                                                                                                                                                                                                                                                                                                                              \\
                        & Time             & 0: 0-15s, 1: 15-30s, 2: 30-45s, 3: 45-60s, 4: \textgreater{}60s, 5: NA                                                                                                                                                                                                                                                                                                                                                                                                                                                                                                                                                                                                                                                                                                                                                                                                                                                                                                                                                                                                                                                                                  \\ \hline
Bar-level               & Chord            & \begin{tabular}[c]{@{}l@{}}0: C:, 1: C:m, 2: C:+, 3: C:dim, 4: C:7, 5: C:maj7, 6: C:m7, \\ 7: C:m7b5, 8: C\#:, 9: C\#:m, 10: C\#:+, 11: C\#:dim, \\ 12: C\#:7, 13: C\#:maj7,  14: C\#:m7, 15: C\#:m7b5, 16: D:, \\ 17: D:m, 18: D:+, 19: D:dim, 20: D:7,  21: D:maj7, \\ 22: D:m7, 23: D:m7b5, 24: Eb:, 25: Eb:m, 26: Eb:+, \\ 27: Eb:dim, 28: Eb:7, 29: Eb:maj7, 30: Eb:m7, 31: Eb:m7b5,\\ 32: E:, 33: E:m, 34: E:+, 35: E:dim, 36: E:7, 37: E:maj7, \\ 38: E:m7,  39: E:m7b5, 40: F:, 41: F:m, 42: F:+, \\ 43: F:dim, 44: F:7, 45: F:maj7,  46: F:m7, 47: F:m7b5, 48: F\#:, \\ 49: F\#:m, 50: F\#:+, 51: F\#:dim, 52: F\#:7, 53: F\#:maj7, \\ 54: F\#:m7, 55: F\#:m7b5, 56: G:, 57: G:m, 58: G:+, 59: G:dim, \\ 60: G:7, 61: G:maj7, 62: G:m7, 63: G:m7b5, 64: Ab:, 65: Ab:m, \\ 66: Ab:+, 67: Ab:dim, 68: Ab:7, 69: Ab:maj7, 70: Ab:m7, \\ 71: Ab:m7b5, 72: A:, 73: A:m, 74: A:+, 75: A:dim, 76: A:7, \\ 77: A:maj7, 78: A:m7,  79: A:m7b5, 80: Bb:, 81: Bb:m, 82: Bb:+,\\  83: Bb:dim, 84: Bb:7,  85: Bb:maj7, 86: Bb:m7, 87: Bb:m7b5, \\ 88: B:, 89: B:m, 90: B:+,  91: B:dim, 92: B:7, 93: B:maj7,\\  94: B:m7, 95: B:m7b5, 96: N.C.\end{tabular} \\ \hline
\end{tabular}
\end{table}

% \begin{table}[htbp]
%     \caption{Accuracy (\%) of each global attribute among MuseCoco, BFT, and Ours.}
%     \centering
% \label{global_acc}
% %\resizebox{\linewidth}{!}{
%     \begin{tabular}{llll}
% \toprule
%  Global attributes & MuseCoco (\%) & BFT (\%) &  Ours (\%) \\ \midrule
%         Instrument        & 96.20    & \textbf{100}  &\textbf{100} \\
%         Pitch Range        & \textbf{61.56 }  & 24.81  & 25.37   \\
%         Rhythm Intensity        &  \textbf{80.47}    & 75.12  & 78.75\\
%         Bar        & 71.80   & \textbf{100}  & \textbf{100}   \\
%         Time Signature        & \textbf{99.14}   & 82.80  & 84.38   \\
%         Key        & 57.42   & \textbf{69.61}  & \textbf{75.00 }  \\
%         Tempo        & 92.71   & \textbf{100}  & 96.88   \\
%         Time        & 65.82   & \textbf{100}  & \textbf{100}   \\
%         Ave.        &78.14    & 81.54  & \textbf{82.55}   \\\bottomrule
% \end{tabular} %}
% \end{table}

\begin{figure}[htbp]
	\centering
	\includegraphics[width=5.5in]
 {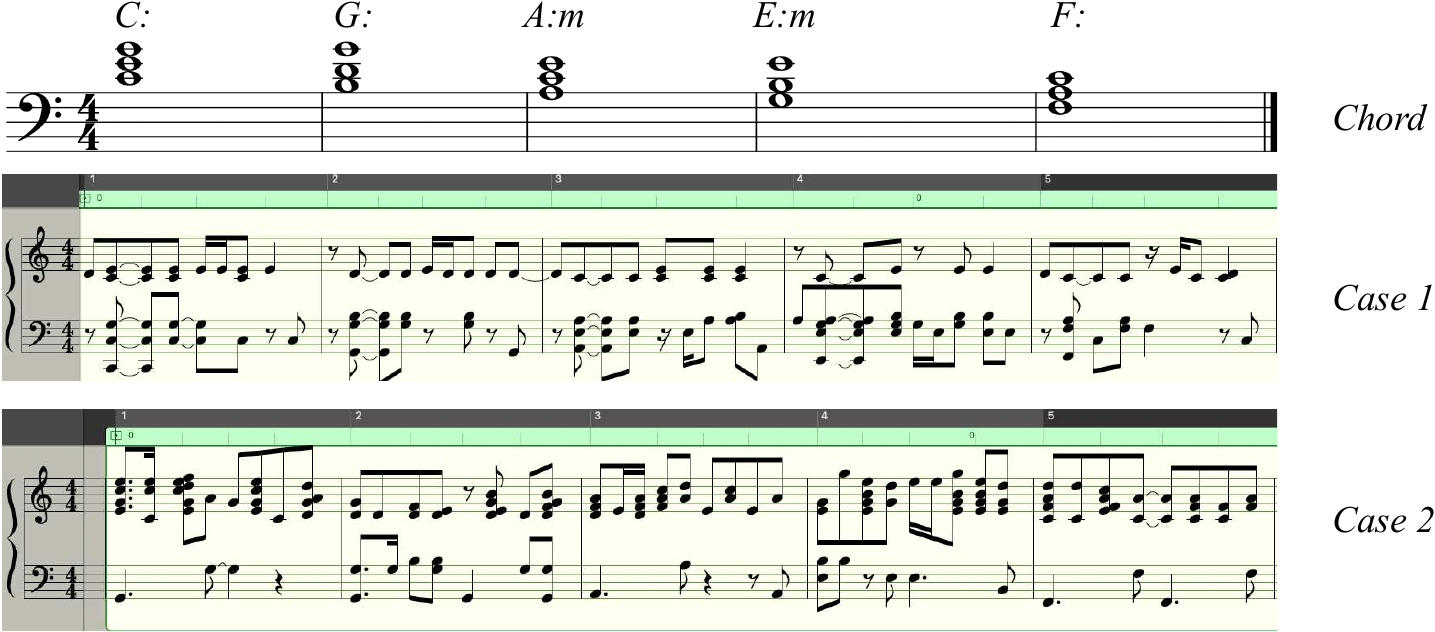}
	\caption{Two generated examples using ``Canon chord progression'' with different global attribute controls.}
	\label{appendix:canon_case}
\end{figure}

\begin{figure}[htbp]
	\centering
	\includegraphics[width=5.5in]
 {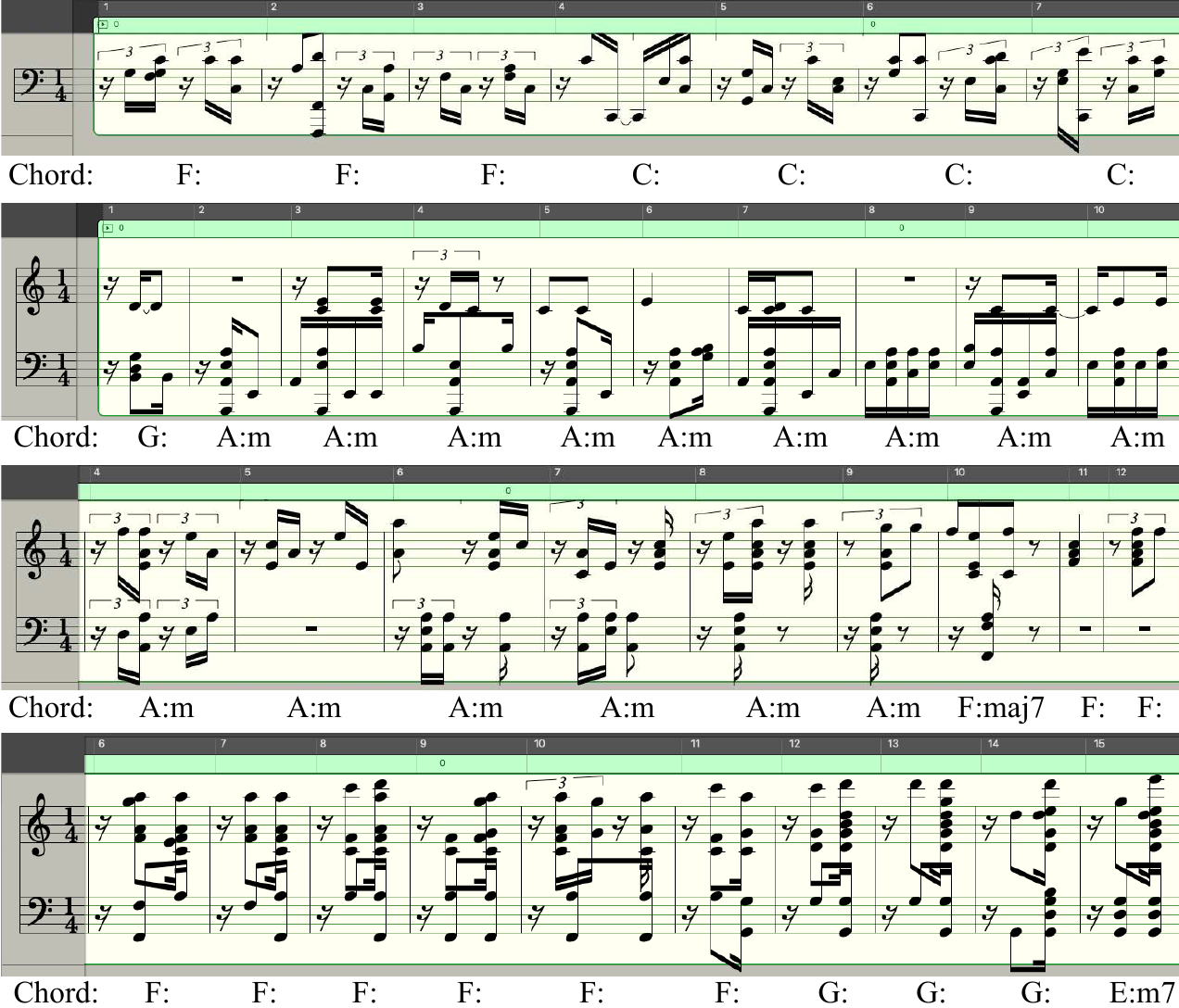}
	\caption{ Some good cases of generated piano rolls, where the chords in the generated music perfectly align with the prompts.}
	\label{good_case}
\end{figure}

\begin{figure}[htbp]
	\centering
	\includegraphics[width=5.5in]
 {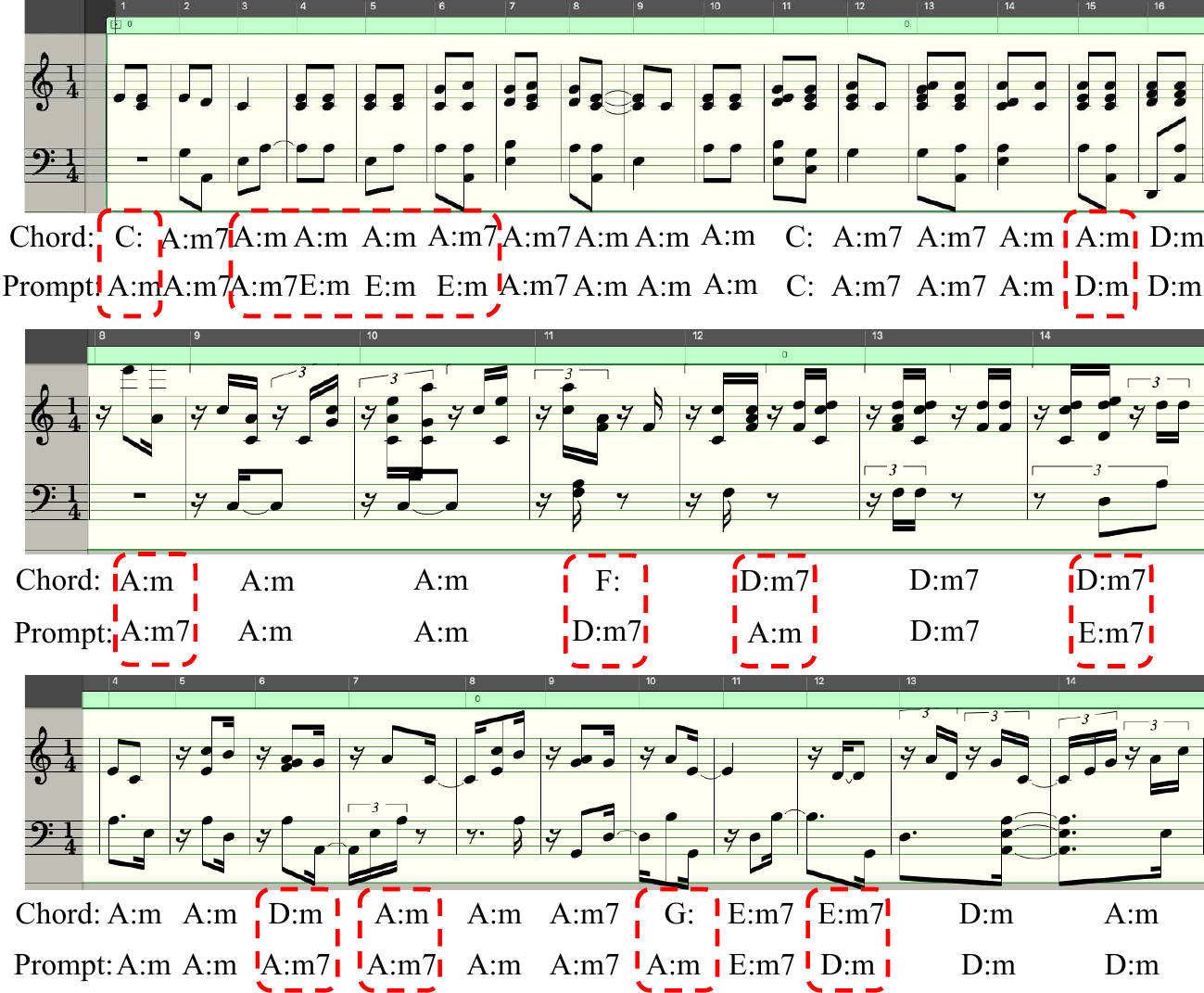}
	\caption{ Examples of failure cases in generated piano rolls, where the chords in the music do not align with the prompts, highlighted by red boxes.}
	\label{fail_case}
\end{figure}

\begin{figure}[htbp]
	\centering
	\includegraphics[width=3.5in]
 {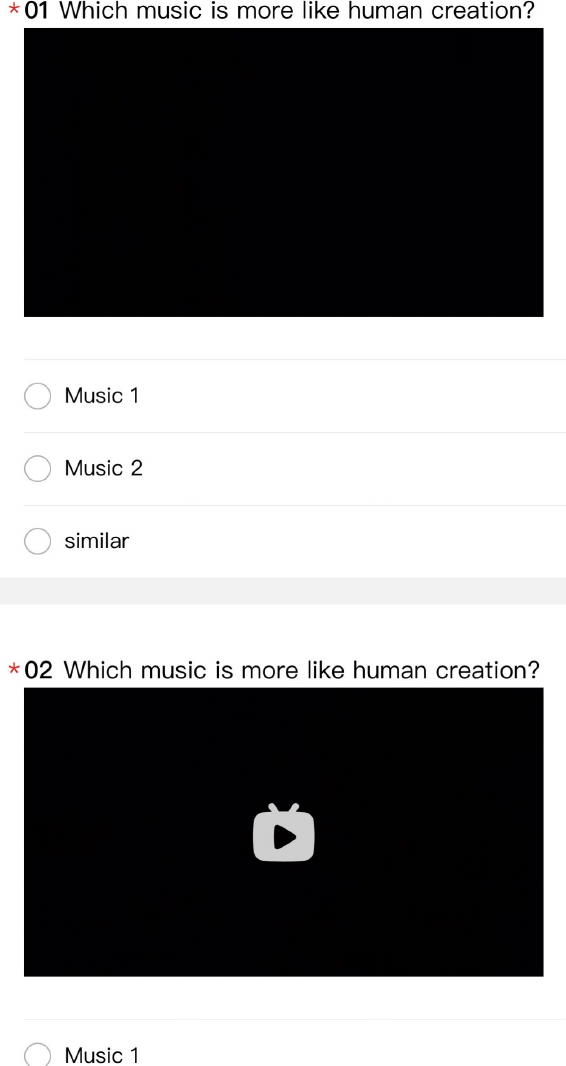}
	\caption{ Screenshot of Musicality Judgment Data Collection: MuseCoco vs. Our Method.}
	\label{appendix_survey}
\end{figure}

% \subsection{Global Attributes Control Accuracy}
% Table~\ref{global_acc} displays the control accuracy for each global attribute on the test dataset. Our method achieves an average global accuracy of 82.55\%, surpassing that of MuseCoco and BFT.

\subsection{Piano Rolls Analysis}
\label{Appendix_pinao_rolls}

\subsubsection{Canon-Style Case Generation}

We conducted experiments using the ``Canon chord progression'', as shown in Figure~\ref{method:example}, to generate new melodies. In addition, we applied different global attributes to generate distinct melodies while maintaining the Canon style. The global attributes for the first case are 4 octaves, intense intensity, moderato tempo, and minor key. For the second case, the attributes are 2 octaves, moderate intensity, fast tempo, and major key. The piano rolls for these two cases are displayed in Figure~\ref{appendix:canon_case}, and the audio can be found in the Supplementary Material. Interestingly, we found that both cases retain the Canon style, with melodies that are comfortable, beautiful, and distinctly Canon-like.

\subsubsection{Good Cases}
We selected some successful cases from our testing where the generated chords perfectly align with the prompts at the bar level, as shown in Figure~\ref{good_case}. This demonstrates the proposed method's strong ability to control each bar's generation with the correct chords.
\subsubsection{Failure Cases}
We also selected some failure cases to analyze and observe how the model failed to generate the correct chords in each bar, as shown in Figure~\ref{fail_case}. In these instances, we found that some prompt chords, such as A:m7, were incorrectly generated as A:m, and D:m7 was incorrectly generated as F. Additionally, the model frequently confused D:m, D:m7, and E:m7. This indicates that the model sometimes struggles to differentiate between similar chords (e.g., A:m: A-C-E vs A:m7: A-C-E-G; D:m7: D-F-A-C vs F: F-A-C).

\subsection{Chord Proportions.}
\label{Chord_Proportions}
\begin{table}[htbp]
\caption{Statistics of chord proportions in the POP909 dataset. \textcolor{blue}{blue} represents the chord and its proportion more than 1\%. Since the selected dataset mainly focuses on popular music, it does not cover all possible chords. However, to ensure the scalability of the method, we uniformly model all chords.}
\label{Appendix_chord_proportion}
\centering
\begin{tabular}{cccccc}
\hline
Chord    & Proportion (\%) & Chord    & Proportion (\%) & Chord                & Proportion (\%)          \\ \hline
\textcolor{blue}{C:}       & \textcolor{blue}{17.11}      & \textcolor{blue}{E:m}      & \textcolor{blue}{6.97}       & Ab:+                 & 0                    \\
C:m      & 0.13       & E:+      & 0          & Ab:dim               & 0.04                 \\
C:+      & 0.08       & E:dim    & 0.05       & Ab:7                 & 0.02                 \\
C:dim    & 0.01       & E:7      & 0.21       & Ab:maj7              & 0.06                 \\
C:7      & 0.17       & E:maj7   & 0.01       & Ab:m7                & 0.01                 \\
\textcolor{blue}{C:maj7}   & \textcolor{blue}{1.62}       & \textcolor{blue}{E:m7}     & \textcolor{blue}{3.24}       & Ab:m7b5              & 0                    \\
C:m7     & 0.05       & E:m7b5   & 0.03       & A:                   & 0.63                 \\
C:m7b5   & 0.01       & \textcolor{blue}{F:}       & \textcolor{blue}{9.94}       & \textcolor{blue}{A:m}                  & \textcolor{blue}{18.31}               \\
C\#:     & 0.38       & F:m      & 0.57       & A:+                  & 0                    \\
C\#:m    & 0.02       & F:+      & 0          & A:dim                & 0.01                 \\
C\#:+    & 0.01       & F:dim    & 0.01       & A:7                  & 0.07                 \\
C\#:dim  & 0.02       & F:7      & 0.02       & A:maj7               & 0.02                 \\
C\#:7    & 0.01       & \textcolor{blue}{F:maj7}   & \textcolor{blue}{2.25}       & \textcolor{blue}{A:m7}                 & \textcolor{blue}{3.91}                 \\
C\#:maj7 & 0.06       & F:m7     & 0.12       & A:m7b5               & 0.01                 \\
C\#:m7   & 0.02       & F:m7b5   & 0          & Bb:                  & 0.52                 \\
C\#:m7b5 & 0.01       & F\#:     & 0.16       & Bb:m                 & 0.23                 \\
\textcolor{blue}{D:}       & \textcolor{blue}{1.18}       & F\#:m    & 0.09       & Bb:+                 & 0                    \\
\textcolor{blue}{D:m}      & \textcolor{blue}{8.03}       & F\#:+    & 0          & Bb:dim               & 0                    \\
D:+      & 0          & F\#:dim  & 0.20       & Bb:7                 & 0.01                 \\
D:dim    & 0.06       & F\#:7    & 0          & Bb:maj7              & 0.07                 \\
D:7      & 0.18       & F\#:maj7 & 0.04       & Bb:m7                & 0.08                 \\
D:maj7   & 0.03       & F\#:m7   & 0.05       & Bb:m7b5              & 0                    \\
\textcolor{blue}{D:m7}     & \textcolor{blue}{3.02}       & F\#:m7b5 & 0.08       & B:                   & 0.07                 \\
D:m7b5   & 0.09       & \textcolor{blue}{G:}       & \textcolor{blue}{14.13}      & B:m                  & 0.21                 \\
Eb:      & 0.16       & G:m      & 0.17       & B:+                  & 0                    \\
Eb:m     & 0.09       & G:+      & 0          & B:dim                & 0.22                 \\
Eb:+     & 0.02       & G:dim    & 0.01       & B:7                  & 0.01                 \\
Eb:dim   & 0.01       & G:7      & 0.98       & B:maj7               & 0                    \\
Eb:7     & 0.01       & G:maj7   & 0.08       & B:m7                 & 0.12                 \\
Eb:maj7  & 0.03       & G:m7     & 0.12       & B:m7b5               & 0.48                 \\
Eb:m7    & 0.05       & G:m7b5   & 0.01       & N.C.                 & 0.37                 \\
Eb:m7b5  & 0.01       & Ab:      & 0.45       &                      &                      \\
\textcolor{blue}{E:}       & \textcolor{blue}{1.84}       & Ab:m     & 0.01       & \multicolumn{1}{l}{} & \multicolumn{1}{l}{} \\ \hline
\end{tabular}
\end{table}
The complete chord distribution is presented in Table~\ref{Appendix_chord_proportion}. As shown, the common chords in the POP909 dataset include C, C:maj7, D, D:m, D:m7, E, E:m, E:m7, F, F:maj7, G, A:m, and A:m7. Some chords, however, never appear in the dataset, such as D:+, E:+, F:+, F:m7b5, F\#:+, F\#:7, G:+, Ab:+, Ab:m7b5, A:+, Bb:+, Bb:dim, Bb:m7b5, B:+, and B:maj7. Given that the selected dataset primarily focuses on popular music, it does not encompass all possible chords. Nonetheless, to ensure the scalability of our method, we uniformly model all chords.
\subsection{Human Evaluation}
Figure~\ref{appendix_survey} shows the voting interface. 

\end{document}